\documentclass[11pt]{article}
\usepackage{fullpage,epsf,amssymb,amsthm,amsfonts,amsmath,latexsym,array,graphicx,extarrows,mathtools,mathstyle,mathrsfs,slashed,subcaption,amsbsy,csquotes,accents}
\usepackage[normalem]{ulem}
\usepackage{authblk}
\usepackage{cite}
\usepackage{color}

\usepackage{float}

\usepackage{hyperref}

\usepackage[export]{adjustbox}

\numberwithin{equation}{section}

\title{\bf{Real-time observables from Euclidean thermal correlation functions}} 

\author[1]{Peter Lowdon\thanks{lowdon@itp.uni-frankfurt.de}}
\author[2,3]{Ralf-Arno Tripolt\thanks{Ralf-Arno.Tripolt@theo.physik.uni-giessen.de}}

\affil[1]{Institut f\"{u}r Theoretische Physik, Johann Wolfgang Goethe-Universit\"{a}t, Max-von-Laue-Str. 1,  60438 Frankfurt am Main, Germany}
\affil[2]{Institut f\"{u}r Theoretische Physik, Justus-Liebig-Universit\"{a}t, Heinrich-Buff-Ring 16, 35392 Giessen, Germany}
\affil[3]{Helmholtz Research Academy Hesse for FAIR (HFHF), Campus Giessen, 35392 Giessen, Germany}

\date{}
\begin{document}

{\let\newpage\relax\maketitle}
\setcounter{page}{1}
\pagestyle{plain}

\begin{abstract}
\noindent
In this work we apply a local quantum field theory approach in order to analyse the connection between real-time observables and Euclidean thermal correlation functions. In particular, using data generated from the functional renormalisation group in the quark-meson model, we demonstrate that in-medium effects can be directly extracted from the spatial momentum dependence of the Euclidean propagators, in contrast to conventional approaches, which rely on the reconstruction from different Matsubara frequencies. As an application, we determine the analytic features that arise from the discrete spectral contribution to the pion correlation function, and calculate the non-perturbative shear viscosity arising from these states.
\end{abstract}

\newpage

\section{Introduction}
\label{intro}

As particles move through thermal media they experience non-trivial effects due to their interactions with constituents in the background. These medium effects give rise to important physical phenomena including the screening of particle masses and the restoration of symmetries. In order to correctly describe these effects one ultimately requires a framework that does not depend on the coupling regime of the system. In Refs.~\cite{Bros:1992ey,Bros:1995he,Bros:1998ua,Bros:1996mw,Bros:2001zs} important steps were taken in this direction by constructing a $T>0$ generalisation of local formulations of quantum field theory (QFT), whose applications over the years have led to numerous foundational insights into vacuum-state systems~\cite{Streater:1989vi,Haag:1992hx,Bogolyubov:1990kw}. For simplicity, this generalised $T>0$ framework focussed on the case of Hermitian scalar fields $\phi(x)$, and established that characteristic features such as the loss Lorentz symmetry can be incorporated by defining a thermal background state $|\Omega_{\beta}\rangle$ at temperature $T=1/\beta$ which is no longer invariant under the full Poincar\'{e} group. Analogously to the vacuum-state case, all of the dynamical properties of the theory are encoded in the thermal correlation functions $\langle \Omega_{\beta}|\phi(x_{1})\cdots\phi(x_{n})|\Omega_{\beta}\rangle$, and these objects are therefore key to unravelling the structure of finite-temperature QFTs. \\  
  
\noindent
Of the techniques that exist to calculate non-perturbative observables, most are either restricted to, or optimised for, the calculation of thermal correlation functions at \textit{imaginary} times. In order to draw consistent conclusions about the physical Minkowski theory using these techniques it is therefore essential to understand the relationship between imaginary and real-time QFTs at non-vanishing temperature. Although Refs.~\cite{Bros:1992ey,Bros:1995he,Bros:1998ua,Bros:1996mw,Bros:2001zs} discussed the general non-perturbative properties satisfied by Euclidean thermal correlation functions, these studies focussed on the formalisation of standard textbook results~\cite{Kapusta:2006pm,Bellac:2011kqa}, such as the periodicity of correlation functions in imaginary time. Further progress was made recently in Ref.~\cite{Lowdon:2022keu}, where it was shown that the Euclidean two-point function is fixed by the spectral properties of the real-time theory. An important consequence of this connection is that the in-medium effects experienced by thermal particle states can be directly extracted from Euclidean correlation function data, avoiding the well-known inverse problem. The aim of this work is to apply the results established in Ref.~\cite{Lowdon:2022keu} to non-perturbative data generated using a functional renormalisation group (FRG) approach~\cite{Berges:2000ew,Polonyi:2001se,Pawlowski:2005xe,Gies:2006wv,Schaefer:2006sr,Kopietz:2010zz,Braun:2011pp,Dupuis:2020fhh}. In particular, we will focus on FRG calculations in the quark-meson model~\cite{Ellwanger:1994wy,Jungnickel:1995fp,Berges:1997eu,Schaefer:2004en, Strodthoff:2011tz,Kamikado:2012bt,Tripolt:2013jra,Tripolt:2014wra,Helmboldt:2014iya,Yokota:2016tip,Eser:2018jqo,Wang:2018osm,Tripolt:2020irx}, which serves as an effective theory for QCD and its chiral properties at low energies. The first goal will be to analyse data in this model at finite temperatures in order to determine the discrete spectral contribution to the pion correlation function. Once obtained, we will use this to understand the analytic characteristics brought about by this contribution, and ultimately calculate the shear viscosity of the corresponding pion states. \\

\noindent
The remainder of this paper is structured as follows: in Sec.~\ref{thermal_corr} we outline the non-perturbative spectral properties satisfied by scalar thermal correlation functions, and summarise the findings of Ref.~\cite{Lowdon:2022keu}, in Sec.~\ref{Euclidean_FRG} we describe and analyse the Euclidean FRG data, and in Sec.~\ref{pion_properties} we use the results of Sec.~\ref{Euclidean_FRG} to investigate the in-medium properties of pions in the quark-meson model. We summarise our main results in Sec.~\ref{concl}.

\section{Thermal correlation functions}
\label{thermal_corr}

In Ref.~\cite{Bros:1996mw} it was demonstrated that the assumptions of local QFT adapted to non-vanishing temperatures impose non-trivial analytic constraints on the structure of thermal correlation functions, in particular the existence of thermal spectral representations. In this section we briefly outline the form of these representations for real scalar fields, together with the results of Ref.~\cite{Lowdon:2022keu}, where it was shown that dissipative thermal properties can be directly extracted from Euclidean two-point functions.

\subsection{Spectral representations}
\label{spec_rep}

By virtue of the locality of the fields\footnote{In this case, by locality we mean: $\left[\phi(x),\phi(y)\right]=0$ for $(x-y)^{2}<0$.}, it turns out that the thermal commutator $C_{\beta}(x-y) =\langle \Omega_{\beta}| \left[\phi(x),\phi(y)\right]|\Omega_{\beta} \rangle$ for real scalar fields has the following spectral representation~\cite{Bros:1992ey}:
\begin{align}
\widetilde{C}_{\beta}(p_{0},\vec{p}) = \int_{0}^{\infty} \! ds \int \! \frac{d^{3}\vec{u}}{(2\pi)^{2}} \ \epsilon(p_{0}) \, \delta\!\left(p^{2}_{0} - (\vec{p}-\vec{u})^{2} - s \right) \widetilde{D}_{\beta}(\vec{u},s),
\label{commutator_rep}
\end{align}
where $\widetilde{D}_{\beta}(\vec{u},s)$ is a thermal spectral density that characterises the interactions with the background medium. In the zero-temperature limit ($\beta\rightarrow \infty$) Eq.~\eqref{commutator_rep} reduces to the vacuum K\"{a}ll\'{e}n-Lehmann representation~\cite{Kallen:1952zz,Lehmann:1954xi}, and hence Eq.~\eqref{commutator_rep} represents the corresponding $T>0$ generalisation. \\

\noindent
Similarly, the retarded $r_{\beta}(x) = i\theta(x^{0})C_{\beta}(x)$ and advanced $a_{\beta}(x) = -i\theta(-x^{0})C_{\beta}(x)$ thermal propagators also possess a spectral representation. In the case that $\widetilde{C}_{\beta}(p)$ vanishes in some energy-momentum region, these propagators are recovered as the boundary values $k_{0}\rightarrow p_{0} \pm i\epsilon$ of a single analytic function $\widetilde{G}_{\beta}(k_{0},\vec{p})$~\cite{Bros:2001zs}, and this representation has the form
\begin{align}
\widetilde{G}_{\beta}(k_{0},\vec{p}) = -\int_{0}^{\infty} \! ds \int \frac{d^{3}\vec{u}}{(2\pi)^{3}}   \, \frac{\widetilde{D}_{\beta}(\vec{u},s)}{k^{2}_{0} - (\vec{p}-\vec{u})^{2} - s }.
\label{propagator_rep}
\end{align}
In Sec.~\ref{damping_corr} we will use the representations in Eqs.~\eqref{commutator_rep} and~\eqref{propagator_rep} in order to explore the properties of pion correlation functions using the FRG data analysed in Sec.~\ref{Euclidean_FRG}.

\subsection{Finite-temperature properties from Euclidean data}
\label{diss}

In Euclidean spacetime the thermal two-point function  $\mathcal{W}_{E}(\tau, \vec{x})$ is $\beta$-periodic, and therefore possesses a series representation 
\begin{align}
\mathcal{W}_{E}(\tau, \vec{x}) = \frac{1}{\beta} \! \sum_{N=-\infty}^{\infty} \! w_{N}(\vec{x}) \, e^{\frac{2\pi i N}{\beta}\tau},
\label{Fourier_series}
\end{align} 
with $w_{N}(\vec{x})$ the corresponding Fourier coefficients. In Ref.~\cite{Lowdon:2022keu} it was established that these coefficients are related to the position space thermal spectral density in the following manner:
\begin{align}
w_{N}(\vec{x}) =\frac{1}{4\pi |\vec{x}|}\int_{0}^{\infty} \! ds \ e^{-|\vec{x}|\sqrt{s+\omega_{N}^{2}}} D_{\beta}(\vec{x},s), 
\label{fourier_coeff_matsu}
\end{align}
where $\omega_{N}= \frac{2\pi N}{\beta}$ are the Matsubara frequencies. Given that a theory contains a single particle state of mass $m$ at $T=0$, a natural assumption is that the thermal spectral density has the decomposition~\cite{Bros:2001zs}
\begin{align}
D_{\beta}(\vec{x},s)= D_{m,\beta}(\vec{x})\, \delta(s-m^{2}) + D_{c, \beta}(\vec{x},s),
\label{decomp}
\end{align} 
where $D_{c, \beta}(\vec{x},s)$ is continuous in the variable $s$ and non-vanishing for $s \geq s_{c}$. The coefficient $D_{m,\beta}(\vec{x})$ has the properties of a damping factor, since by virtue of the structure of Eq.~\eqref{commutator_rep} its non-triviality causes the zero-temperature mass pole to be screened. Combining Eqs.~\eqref{fourier_coeff_matsu} and~\eqref{decomp} it was further shown in Ref.~\cite{Lowdon:2022keu} that if the gap between $m$ and the continuum onset $s_{c}$ is sufficiently large, then the damping factor can be estimated from the $N=0$ Fourier coefficient
\begin{align}
D_{m,\beta}(\vec{x}) \sim 4\pi |\vec{x}|\, e^{|\vec{x}|m} w_{0}(\vec{x}).
\label{Fourier_coeff_rel}
\end{align} 
Although one can attempt to construct analogous relations to Eq.~\eqref{Fourier_coeff_rel} for $N>0$, the dominance of the damping factor contribution to Eq.~\eqref{fourier_coeff_matsu} is no longer guaranteed, since the suppression of $D_{c, \beta}(\vec{x},s)$ is increasingly diminished for larger values of $N$, and for higher temperatures. The $N=0$ Fourier coefficient $w_{0}(\vec{x})$ is therefore optimal for extracting the behaviour of the damping factor. For the purpose of the analysis in this study we are interested in momentum space Euclidean data, in particular the propagator $\widetilde{G}_{\beta}(k_{0},\vec{p})$ at different Matsubara frequencies $k_{0}=i\omega_{N}$. In this case, $w_{0}(\vec{x})$ can be computed via the relation     
\begin{align}
w_{0}(\vec{x}) =    \frac{1}{2\pi^{2}|\vec{x}|}\int_{0}^{\infty} \! d|\vec{p}|\ |\vec{p}| \sin\!\left(|\vec{p}||\vec{x}|\right) \, \widetilde{G}_{\beta}(0,|\vec{p}|),
\label{Fourier_coeff_dampingp}
\end{align}    
where here we make use of the fact that the propagator depends only on the absolute value of $\vec{p}$, which follows from the assumption of rotational invariance.

\section{Euclidean FRG data}
\label{Euclidean_FRG}

The main objective of this study was to use non-perturbative FRG data in order to test the analytic results outlined in Sec.~\ref{diss}, in particular Eq.~\eqref{Fourier_coeff_rel}. In this section we describe the FRG data and the corresponding analysis strategy.

\subsection{Data description}
\label{description}

\begin{figure}[H]
\centering
\includegraphics[width=0.55\columnwidth]{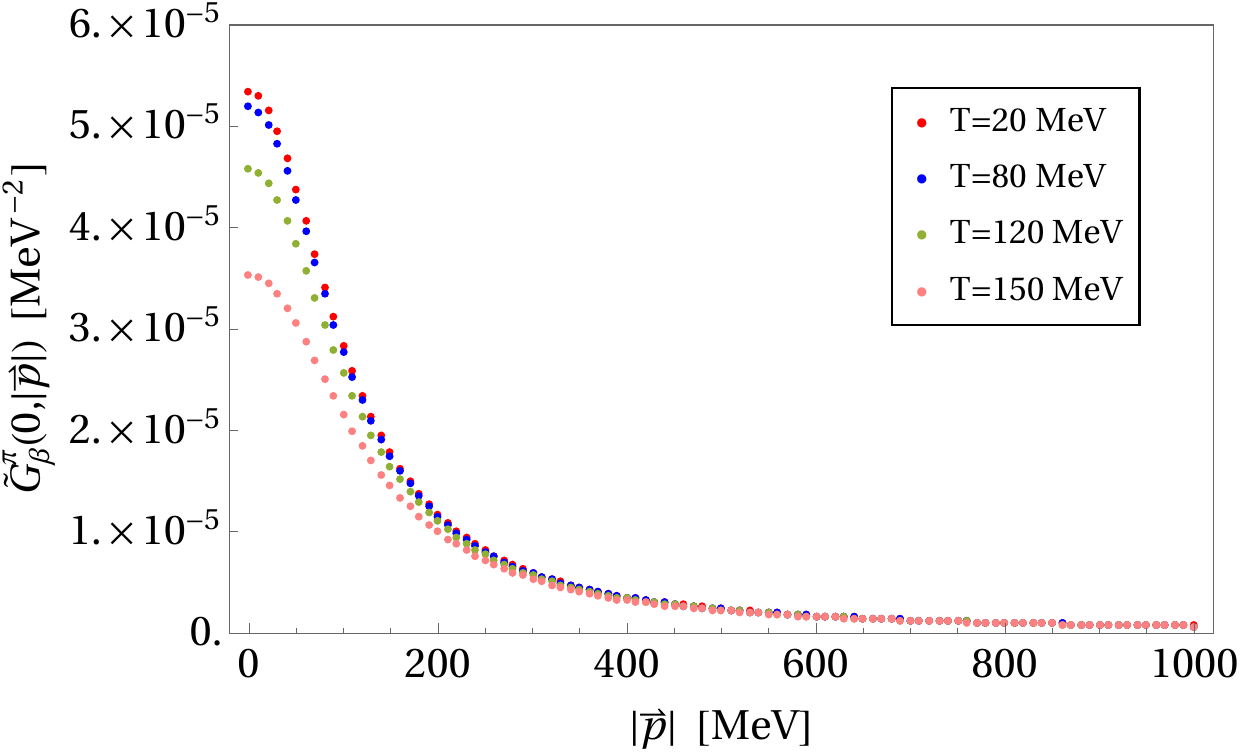}
\caption{Plot of the $\widetilde{G}_{\beta}^{\pi}(k_{0}=0,|\vec{p}|)$ data as a function of $|\vec{p}|$ for different temperatures.}
\label{G_plot}
\end{figure}

For the purpose of this study we applied an FRG framework based on the theoretical setup in Refs.~\cite{Tripolt:2013jra,Tripolt:2014wra} to generate Euclidean data of the pion propagator in a similar manner to Ref.~\cite{Tripolt:2018xeo}. In particular, in order to compute $w_{0}(\vec{x})$ via Eq.~\eqref{Fourier_coeff_dampingp} we calculated the pion propagator at zero Matsubara frequency and non-vanishing spatial momenta at different temperatures in the range $[1,150] \, \text{MeV}$. A selection of these data are plotted in Fig.~\ref{G_plot}. More details regarding the precise theoretical setup and employed parameter configuration can be found in Refs.~\cite{Tripolt:2013jra,Tripolt:2014wra}. Since the FRG data is only meaningful up to an ultraviolet cutoff scale, which in the present case is given by $\Lambda=1 \, \text{GeV}$, we used an extrapolation for larger momenta in order to compute the integral in Eq.~\eqref{Fourier_coeff_dampingp}. This extrapolation was based on the general analytic result that $\widetilde{G}_{\beta}(0,|\vec{p}|) \sim 1/|\vec{p}|^{2}$ in the large $|\vec{p}|$ regime. More specifically, depending on the value of $|\vec{x}|$, the integral in Eq.~\eqref{Fourier_coeff_dampingp} was split into two parts; a low-momentum part, where the integral was evaluated in the range $[0,p_{\text{max}}]$ with $p_{\text{max}}$ containing the largest number of full cycles of the $\sin$ function in the integrand of Eq.~\eqref{Fourier_coeff_dampingp} based on the propagator data, and a high-momentum part, evaluated in the range $[p_{\text{max}},\infty)$ using a continuous extrapolation based on the expected $1/|\vec{p}|^{2}$ asymptotic form.
 
\subsection{Data analysis}
\label{analysis}

\begin{figure}[H]
\centering
\includegraphics[width=0.58\columnwidth]{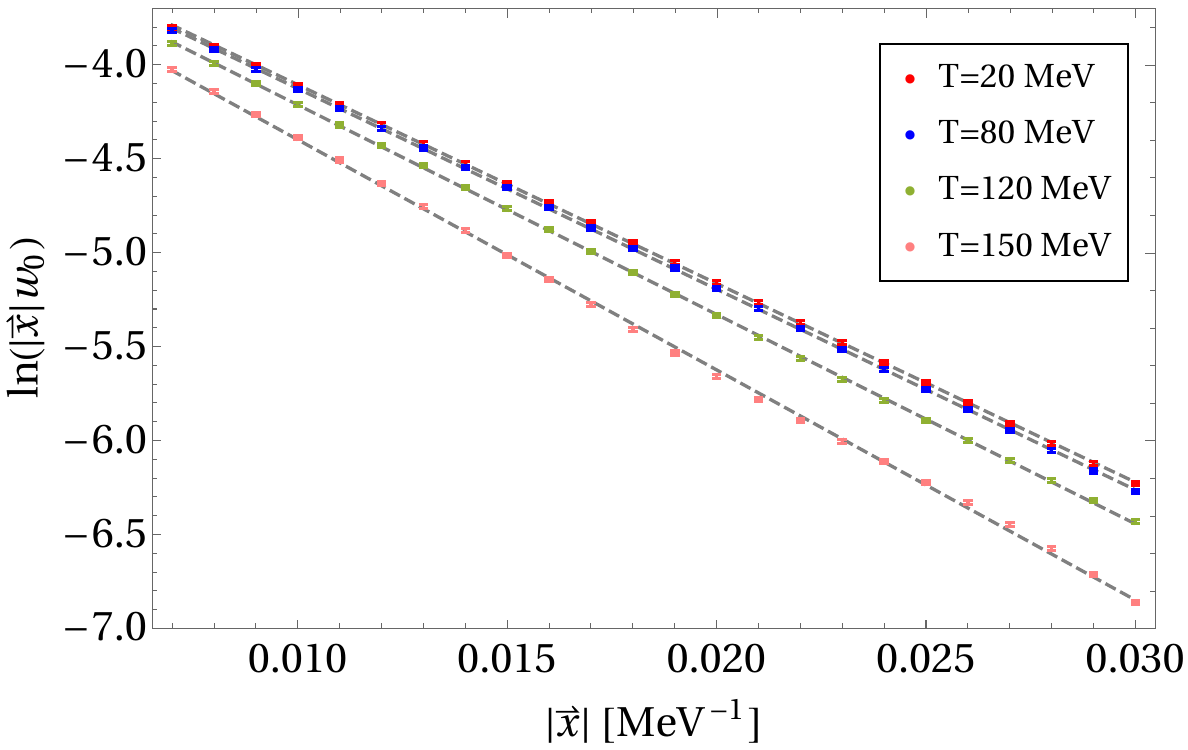}
\caption{Plot of the $\ln\left(|\vec{x}|w_{0}(\vec{x}) \right)$ data together with their respective errors and best linear fits (grey dashed lines) for different temperatures.}
\label{logdata_plot}
\end{figure}

In this section we outline the strategy we adopted to analyse the pion FRG data described in Sec.~\ref{description}. The goal of this analysis was to test whether Eq.~\eqref{Fourier_coeff_rel} could be used to extract a meaningful damping factor expression from the data, and if so, what form this takes. Based on a qualitative inspection we found that $|\vec{x}|w_{0}(\vec{x})$ possessed an exponential-type behaviour, and so we performed fits of the linear ansatz $f(\vec{x})=A -B |\vec{x}|$ to the $\ln\left(|\vec{x}|w_{0}(\vec{x})  \right)$ data at each value of $T$. The $|\vec{x}|$-range of these fits was chosen so as to minimise the numerical errors from the FRG computation. To assess the overall quality of the fits we calculated the $\chi^{2}/\text{d.o.f.}$ under the assumption of a 1\% systematic error, which was estimated by varying the choice of extrapolation used to compute $w_{0}(\vec{x})$. Although this gave an approximate estimate of the effect of possible deviations at intermediate momenta, we found that increasing the total systematic error by 1\%-2\% did not lead to a significant deterioration in the quality of the fits. Overall, we found that these data were well described by the linear ansatz function, obtaining $\chi^{2}/\text{d.o.f.} \lesssim 1$ in the range $T \in [1,125] \, \text{MeV}$, and slightly larger values for $T>125 \, \text{MeV}$. A selection of these data, together with their best linear fits, are plotted in Fig.~\ref{logdata_plot}.    
   
Overall, the result of these fits demonstrated that the data was consistent with the existence of a pion damping factor with the following functional form:    
\begin{align}
D_{\pi,\beta}(\vec{x}) = \alpha \,  e^{-|\vec{x}|\gamma},
\label{damping_x}
\end{align}
where the parameters are related to the fitted quantities by: $\alpha= 4\pi e^{A}$, $\gamma=B-m_{\pi}$. The zero-temperature pion mass was taken to be $m_{\pi}=105.769 \, \text{MeV}$, which was estimated from the zero-momentum intercept of the pion propagator data at the lowest available temperature ($1\, \text{MeV}$), where thermal effects are negligible and the pion pole dominates, together with the analytic bound\footnote{This bound arises from the general fact that damping factors are tempered distributions, and therefore can only grow at most polynomially in $|\vec{x}|$.} $\gamma \geq 0$. As one can see in Fig.~\eqref{logdata_plot}, both the intercept and slope of the data display a clear temperature dependence, which implies that $\{\alpha,\gamma\}$ and hence $D_{\pi,\beta}(\vec{x})$ are functions of $T$, as one would expect.

\section{Real-time observables}
\label{pion_properties}

Based on the analysis in Sec.~\ref{analysis} the FRG data for the pion propagator is well-described in terms of the damping factor of Eq.~\eqref{damping_x}. In this section we will explore the implications of such a damping factor on the in-medium properties of pions in the quark-meson model. 

\subsection{Pion damping factor and thermal correlation functions}
\label{damping_corr}

Upon taking the Fourier transform of Eq.~\eqref{damping_x} one finds that the pion damping factor has the momentum space structure 
\begin{align}
\widetilde{D}_{\pi,\beta}(\vec{u}) = \frac{8\pi \alpha \gamma}{(|\vec{u}|^{2}+\gamma^{2})^{2}}.
\label{damping_p}
\end{align}
As outlined in Sec.~\ref{spec_rep}, the behaviour of the thermal commutator and propagator are both fixed by the thermal spectral density. Therefore, by setting $\widetilde{D}_{\beta}(\vec{u},s) =  \widetilde{D}_{\pi,\beta}(\vec{u})\delta(s-m_{\pi}^{2})$ in Eq.~\eqref{commutator_rep} one can determine the discrete spectral contribution to the pion commutator. In light of the analysis in Ref.~\cite{Bros:2001zs} this corresponds to the commutator of pion states at \textit{asymptotic} times ($x_{0}\rightarrow \pm \infty$), since continuum contributions are suppressed in this regime. The resulting commutator has the form 
\begin{align}
\widetilde{C}_{\beta}^{\pi}(p_{0},\vec{p})=  \epsilon(p_{0})\theta(p_{0}^{2}-m_{\pi}^{2}) \, \frac{\alpha}{|\vec{p}|} \left[ \frac{\gamma}{ \left(|\vec{p}|-\sqrt{p_{0}^{2}-m_{\pi}^{2}} \right)^{2} + \gamma^{2}   } - \frac{\gamma}{ \left(|\vec{p}|+\sqrt{p_{0}^{2}-m_{\pi}^{2}} \right)^{2} + \gamma^{2}   }     \right].
\label{commutator_pion}
\end{align}
In the limit $\gamma\rightarrow 0$ Eq.~\eqref{commutator_pion} reduces to the zero-temperature result $2\pi\alpha \, \epsilon(p_{0}) \delta(p^{2}-m_{\pi}^{2})$, and hence $\gamma$ has the interpretation of a thermal width. Although Eq.~\eqref{commutator_pion} is similar in structure to the Lorentzian ansatz often proposed in the literature\footnote{Further discussion of the Lorentzian ansatz, and in particular its applications to QCD, can be found in Refs.~\cite{Peshier:2004bv,Peshier:2005pp}.} this expression has distinct characteristics, including the existence of a sharp energy cutoff at the zero-temperature particle mass $m_{\pi}$. It is interesting to note that this characteristic is also seen in other non-perturbative models~\cite{Bros:2001zs}. \\  

\begin{figure}[H]
\centering
\includegraphics[width=0.6\columnwidth]{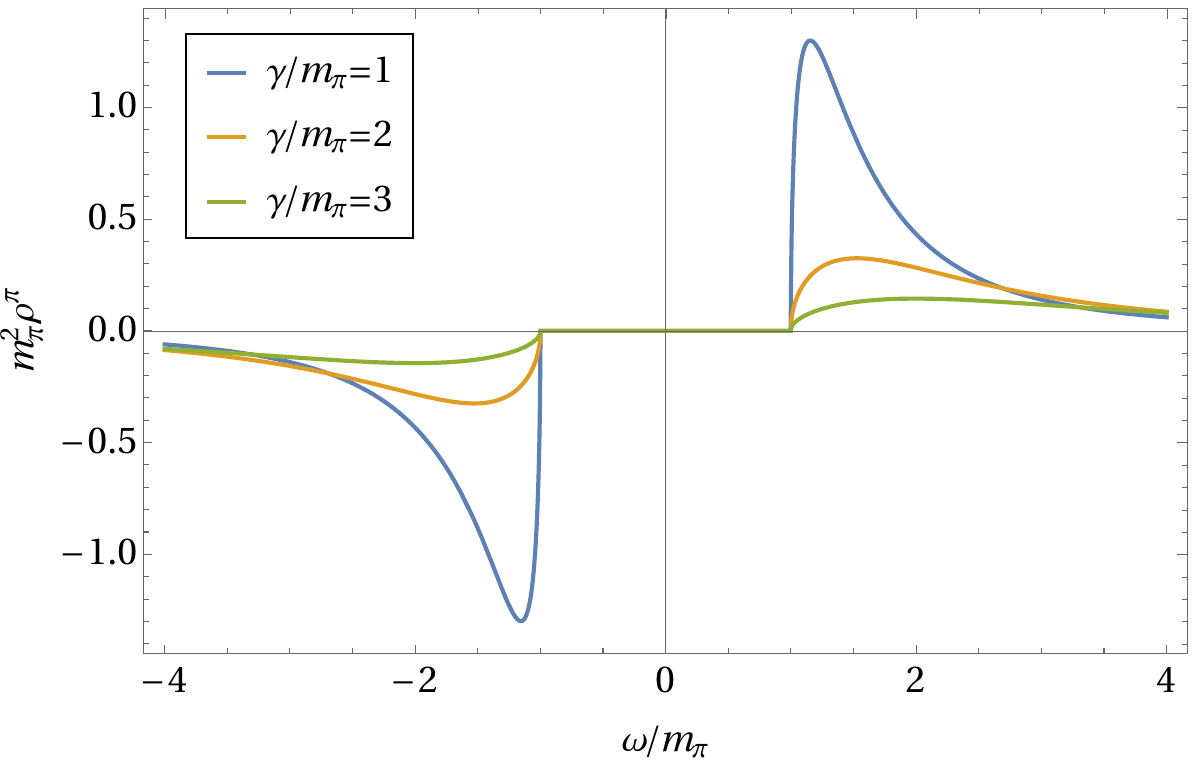}
\caption{Plot of $m_{\pi}^{2}\rho^{\pi}(\omega)$ for different values of $\gamma/m_{\pi}$.}
\label{rho_plot}
\end{figure}

\noindent
Taking the limit $\vec{p}\rightarrow 0$, the resulting spectral function $\rho^{\pi}(\omega)=\widetilde{C}_{\beta}^{\pi}(\omega,\vec{p}=0)$ is given by
\begin{align}                                                   
\rho^{\pi}(\omega)=   \epsilon(\omega)\theta(\omega^{2}-m_{\pi}^{2}) \frac{4\alpha \,  \gamma \sqrt{\omega^{2}-m_{\pi}^{2}}}{(\omega^{2}-m_{\pi}^{2}+\gamma^{2})^{2}}.
\label{spectralF_pion}
\end{align}
In order to get a sense of the structure of Eq.~\eqref{spectralF_pion} the form of the dimensionless expression $m_{\pi}^{2}\rho^{\pi}(\omega)$ is plotted in Fig.~\ref{rho_plot} for different values of the rescaled width $\gamma/m_{\pi}$.  
One finds that the location of the peaks in Fig.~\ref{rho_plot} are at the points $\omega = \pm \sqrt{m_{\pi}^{2}+\frac{\gamma^{2}}{3}}$. Physically, this implies that the zero-temperature mass of the asymptotic pion states is screened as the thermal width $\gamma$ increases, and that this becomes significant in regimes where $\gamma$ is comparable to $m_{\pi}$. \\

\noindent
Since $\widetilde{C}_{\beta}^{\pi}(p)$ vanishes in the region $|p_{0}|<m_{\pi}$, it follows from the discussion in Sec.~\ref{spec_rep} that the corresponding retarded and advanced propagators are the boundary values of a single analytic function $\widetilde{G}_{\beta}^{\pi}(k_{0},\vec{p})$. Performing a similar calculation as with the commutator, and applying Eq.~\eqref{propagator_rep}, one finds that
\begin{align}
\widetilde{G}_{\beta}^{\pi}(k_{0},\vec{p}) =\frac{\alpha}{|\vec{p}|^{2}-k_{0}^{2}+m_{\pi}^{2}+\gamma^{2}+2\gamma\sqrt{m_{\pi}^{2}-k_{0}^{2}}}.
\label{propagator_pion}
\end{align}
As in the case of a Lorentzian-type commutator, the resulting propagator in Eq.~\eqref{propagator_pion} has complex poles in $k_{0}$, which in this case are located at the points
\begin{align}
k_{0} =  -\sqrt{|\vec{p}|^{2}+m_{\pi}^{2} -\gamma^{2} \pm 2 i |\vec{p}|\gamma}, \quad \sqrt{|\vec{p}|^{2}+m_{\pi}^{2} -\gamma^{2} \pm 2 i |\vec{p}|\gamma}.
\label{poles}
\end{align}
A fundamental difference to the Lorentzian case though is that at $\vec{p}=0$ the poles in Eq.~\eqref{poles} are either purely real or imaginary, depending on the relative size of $m_{\pi}$ and $\gamma$.

\subsection{Shear viscosity}

Damping factors also play an essential role in the calculation of real-time observables such as transport coefficients. In particular, in Ref.~\cite{Lowdon:2021ehf} it was demonstrated that the shear viscosity arising from scalar thermal asymptotic states $\eta_{0}$ can be computed explicitly from the damping factors describing those states. In this section we will use Eq.~\eqref{damping_p} to derive the analytic form of $\eta_{0}$ for pions in the quark-meson model, and apply this together with the numerically extracted FRG parameter values to explicitly calculate the temperature dependence of the shear viscosity.

\subsubsection{Analytic form}

\begin{figure}[H]
\centering 
\includegraphics[width=0.6\columnwidth]{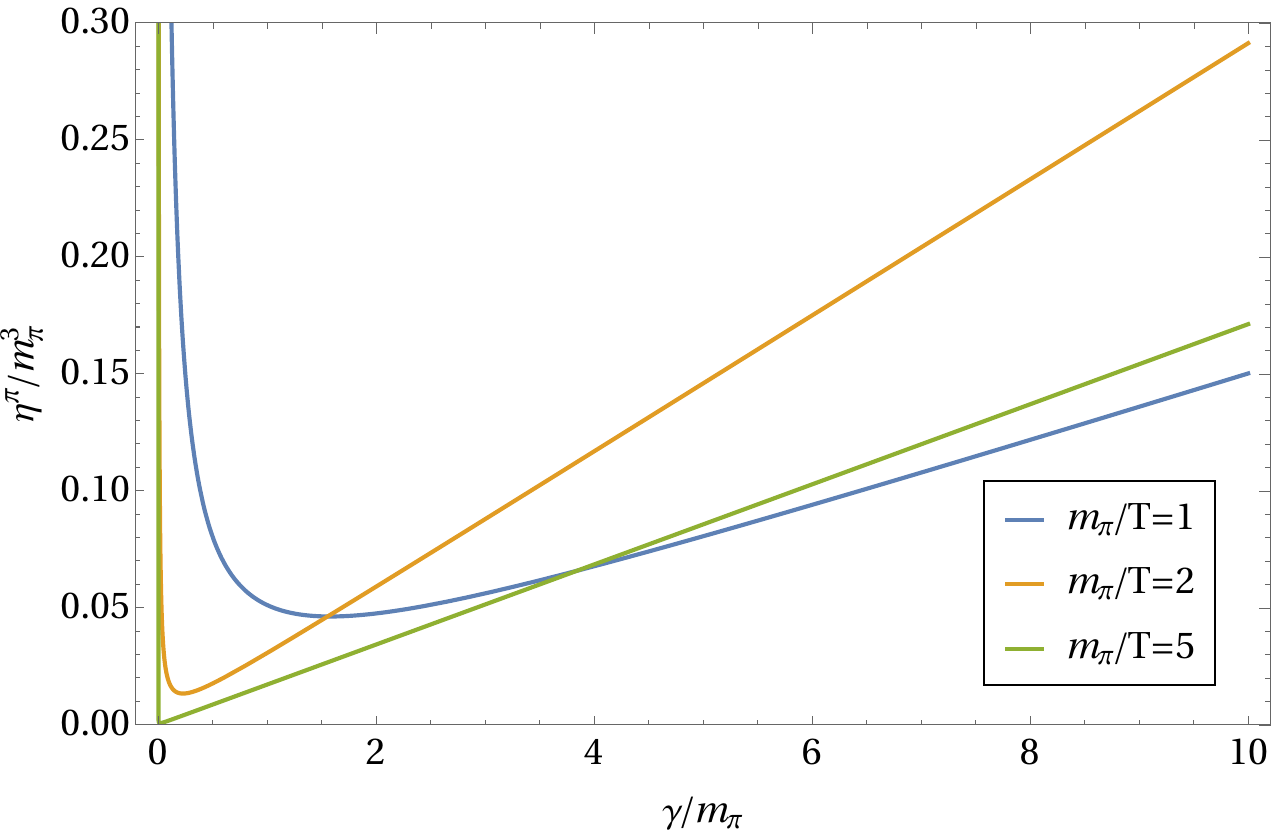}
\caption{Plot of the $\gamma/m_{\pi}$ dependence of $\eta_{0}^{\pi}/m_{\pi}^{3}$ for $\alpha=1$ and different values of $m_{\pi}/T$.}
\label{shear_width}
\end{figure}

Due to the spectral representation in Eq.~\eqref{commutator_rep} it turns out that $\eta_{0}$ has the general form~\cite{Lowdon:2021ehf}  
\begin{align}
\eta_{0} &=  \frac{T^{5}}{240\pi^{5}} \int_{0}^{\infty} \!\! ds \int_{0}^{\infty} \!\! dt\int_{0}^{\infty} d|\vec{u}| \int_{0}^{\infty} d|\vec{v}|  \, |\vec{u}||\vec{v}| \, \widetilde{D}_{\beta}(\vec{u},s) \, \widetilde{D}_{\beta}(\vec{v},t) \nonumber \\[0.5em]
& \hspace{2mm} \times \Bigg[  4\left[1+\epsilon(|\vec{u}|-|\vec{v}|)  \right]\left\{\frac{|\vec{v}|}{T} \, \mathcal{I}_{3}\!\left( \! \frac{\sqrt{t}}{T}, \, 0, \infty  \! \right)  +  \frac{|\vec{v}|^{3}}{T^{3}} \, \mathcal{I}_{1}\!\left( \! \frac{\sqrt{t}}{T}, \, 0, \infty \!\right) \right\}   \nonumber \\[0.5em]
& \hspace{4mm} +   \left\{  \mathcal{I}_{4}\!\left( \! \frac{\sqrt{t}}{T},\frac{|\vec{v}|}{T},\frac{s-t +(|\vec{u}|+|\vec{v}|)^{2}}{2(|\vec{u}|+|\vec{v}|)T}\right) + \epsilon(|\vec{u}|-|\vec{v}|)\, \mathcal{I}_{4}\!\left( \! \frac{\sqrt{t}}{T},\frac{|\vec{v}|}{T},\frac{s-t+(|\vec{v}|-|\vec{u}|)^{2}}{2(|\vec{v}|-|\vec{u}|)T}\right) \right\} \Bigg],
\label{shear_general_st}  
\end{align} 
where $\mathcal{I}_{N}(R,a,b)$ are the class of positive-valued Bose-Einstein distribution-like integrals
\begin{align}
&\mathcal{I}_{N}(R,a,b) =   \int_{0}^{b} \! dQ \frac{(Q-a)^{N}}{ e^{\sqrt{Q^{2}+R^{2}}} -1}.  \label{dimless_int} 
\end{align}
\ \\
\noindent
Using Eq.~\eqref{damping_p}, one can set $\widetilde{D}_{\beta}(\vec{u},s) =  \widetilde{D}_{\pi,\beta}(\vec{u})\delta(s-m_{\pi}^{2})$ in Eq.~\eqref{shear_general_st} in order to calculate the corresponding shear viscosity of the asymptotic pion states $\eta_{0}^{\pi}$. Despite the complexity of Eq.~\eqref{shear_general_st}, ultimately one finds that this reduces to the following simple expression:
\begin{align}
\eta_{0}^{\pi} = \frac{T^{4}\alpha^{2}}{15\pi^{2}\gamma} \, \mathcal{I}_{3}\!\left(  \frac{m_{\pi}}{T}, \, 0, \infty   \right) + \frac{T^{2}\alpha^{2}\gamma}{6\pi^{2}} \, \mathcal{I}_{1}\!\left( \frac{m_{\pi}}{T}, \, 0, \infty \right). 
\label{shear_pion}
\end{align}
In Fig.~\ref{shear_width} we plot the dimensionless quantity $\eta_{0}^{\pi}/m_{\pi}^{3}$ as a function of the rescaled width $\gamma/m_{\pi}$ for different values of $m_{\pi}/T$ and fixed $\alpha$.    
Equation~\eqref{shear_pion} possesses some characteristic features, including the existence of a finite massless limit\footnote{This is fundamentally different to what was found in the calculation of the shear viscosity in $\phi^{4}$ theory, where it was shown that $\eta_{0}$ diverges logarithmically with $m$~\cite{Lowdon:2021ehf}.}, positivity, and a divergent growth for both small and large values of $\gamma$, as shown in Fig~\ref{shear_width}.  

\subsubsection{Numerical computation}

\begin{figure}[H]
\centering
\includegraphics[width=0.65\columnwidth]{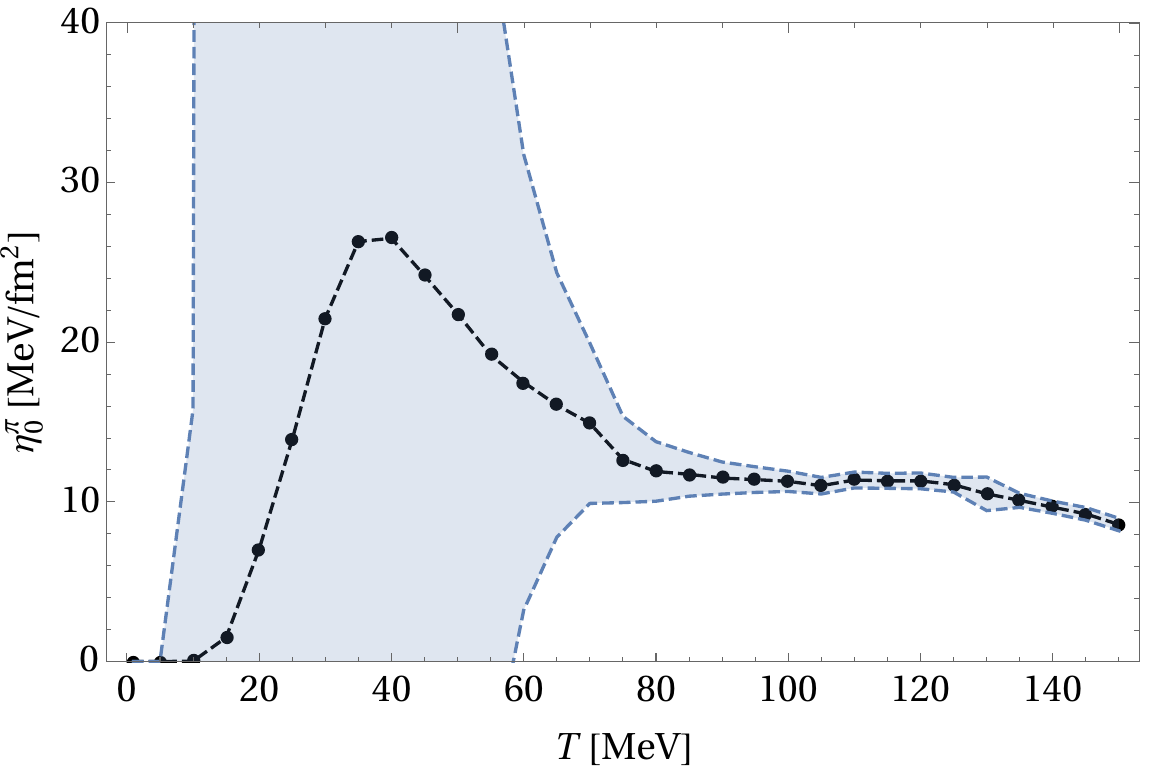}
\caption{Numerical calculation of $\eta_{0}^{\pi}$ as a function of $T$ based on the best-fit parameters extracted from the FRG data. The black points are the central values of $\eta_{0}^{\pi}$, and the blue band signifies the 1$\sigma$ uncertainty.}
\label{shear_plot}
\end{figure}

Now that we have the analytic form of $\eta_{0}^{\pi}$ we can use the $\{\alpha,\gamma\}$ values from the FRG analysis in Sec.~\ref{analysis} to determine $\eta_{0}^{\pi}$. In Fig.~\ref{shear_plot} we plot the resulting temperature dependence of $\eta_{0}^{\pi}$, including the uncertainties estimated from the quality of the numerical fits.     
One can see from Fig.~\ref{shear_plot} that there is significant uncertainty in the range $T\in[1,65] \, \text{MeV}$. This is due to the fact that the $1/\gamma$ component of $\eta_{0}^{\pi}$ dominates at lower values of $T$, and therefore small variations in $\gamma$ can lead to large errors. The 1$\sigma$ uncertainty band in the plot is based on the assumption that the fit parameters are Gaussian distributed, and in the case of $\gamma$ the errors at small $T$ are truncated in order to construct an asymmetric confidence interval, which is necessary since $\gamma$ must be non-negative. However, this procedure results in large relative errors for $\gamma$, and hence $\eta_{0}^{\pi}$, at small $T$. A reduction in the uncertainty of $\eta_{0}^{\pi}$ for $T\in[1,65] \, \text{MeV}$ is certainly achievable with a more sophisticated error analysis and additional data, but we leave this to a future work. In the region where $T$ approaches the lowest values, both $\eta_{0}^{\pi}$ and its associated uncertainty tend to zero because the $\mathcal{I}_{3}$ coefficient in Eq.~\eqref{shear_pion} decays rapidly for vanishing temperature. \\

\noindent
The pion shear viscosity has been computed numerous times in the literature using a variety of different methods such as chiral perturbation theory (ChPT)~\cite{Chen:2006iga,Fernandez-Fraile:2006kxe,Fernandez-Fraile:2009eug,Lang:2012tt} and kinetic theory~\cite{Davesne:1995ms,Dobado:2003wr,Heckmann:2012wqa,Mitra:2012jq}. In particular, comparing Fig.~\ref{shear_plot} with the results of ChPT~\cite{Chen:2006iga,Fernandez-Fraile:2006kxe,Fernandez-Fraile:2009eug,Lang:2012tt} one sees several similar characteristics, including the overall magnitude\footnote{Following the approach of Ref.~\cite{Lang:2012tt} in Fig.~\ref{shear_plot} we plot $\eta_{0}^{\pi}$ in units of $\text{MeV}/\text{fm}^{2}$. An important point to note is that the FRG data does not distinguish between the different pion fields, and so to compare Fig.~\ref{shear_plot} with chiral perturbation theory results one must introduce an overall multiplicative factor of three.} of $\eta_{0}^{\pi}$, the appearance of a peak-like behaviour at low $T$, and the vanishing of $\eta_{0}^{\pi}$ in the $T\rightarrow 0$ limit. From a phenomenological perspective the dimensionless specific shear viscosity $\eta/s$ is also a quantity of particular interest. However, in order to consistently determine $\eta_{0}^{\pi}/s_{\pi}$ in our case this would require a separate calculation of the pion entropy density $s_{\pi}$, which is non trivial. We therefore leave this to a future work.

\section{Conclusions} 
\label{concl}

Euclidean thermal correlation functions play an important role in the characterisation of in-medium effects in finite-temperature QFTs. In Ref.~\cite{Lowdon:2022keu} it was demonstrated for scalar QFTs that the dissipative characteristics of thermal particle states can be directly extracted from these quantities. In particular, the damping factors associated with specific states are related to the form of the Euclidean propagator at zero Matsubara frequency. In this study we apply these findings to non-perturbative Euclidean data of the quark-meson model generated using an FRG approach. For the pion propagator we demonstrate that these data are consistent with the relations derived in Ref.~\cite{Lowdon:2022keu}, and are able to extract the explicit form of the pion damping factor. With this damping factor we determine the distinctive analytic contributions to the thermal correlation functions, and calculate the non-perturbative shear viscosity arising from the asymptotic thermal pion states. This study represents an important step both in the understanding and calculation of non-perturbative real-time observables, and ultimately could provide new insights into fundamental theories such as QCD.

\section*{Acknowledgements}
The authors would like to thank Jan Pawlowski and Lorenz von Smekal for useful discussions and input. This work is supported by the Deutsche Forschungsgemeinschaft (DFG, German Research Foundation) through the Collaborative Research Center CRC-TR 211 ``Strong-interaction matter under extreme conditions'' -- Project No. 315477589-TRR 211.

\bibliographystyle{JHEP}

\bibliography{refs}

\end{document}